\documentclass{ptapap}
\usepackage{amsmath,amssymb}
\author{Alexandre David-Uraz}[UD]
\author{Coralie Neiner}[PSL]
\author{James Sikora}[Bishop]
\author{James Barron}[Queens,RMC]
\author{Dominic M. Bowman}[KULeuven]
\author{P\i{}nar Cerraho\u{g}lu}[UD]
\author{David H. Cohen}[Swarthmore]
\author{Christiana Erba}[UD]
\author{Oleksandr Kobzar}[Moncton]
\author{Oleg Kochukhov}[Uppsala]
\author{V\'{e}ronique Petit}[UD]
\author{Matthew E. Shultz}[UD]
\author{Asif ud-Doula}[PSU]
\author{Gregg A. Wade}[RMC]
\author{the MOBSTER Collaboration}[]
\affil[UD]{Department of Physics and Astronomy, University of Delaware, Newark, DE 19716, USA}
\affil[PSL]{LESIA, Paris Observatory, PSL University, CNRS, Sorbonne University, Université de Paris, 5 place Jules Janssen, 92195 Meudon, France}
\affil[Bishop]{Physics and Astronomy Department, Bishop's University, Sherbrooke, QC J1M 1Z7, Canada}
\affil[Queens]{Department of Physics, Engineering Physics \& Astronomy, Queen's University, 64 Bader Lane, Kingston, ON K7L 3N6, Canada}
\affil[KULeuven]{Institute of Astronomy, KU Leuven, Celestijnenlaan 200D, B-3001 Leuven, Belgium}
\affil[Swarthmore]{Department of Physics and Astronomy, Swarthmore College, Swarthmore, PA 19081, USA}
\affil[Moncton]{D\'{e}partement de Physique et d'Astronomie, Universit\'{e} de Moncton, Moncton, NB E1A 3E9, Canada}
\affil[Uppsala]{Department of Physics and Astronomy, Uppsala University, Box 516, 75120, Uppsala, Sweden}
\affil[PSU]{Penn State Scranton, 120 Ridge View Drive, Dunmore, PA 18512, USA}
\affil[RMC]{Department of Physics and Space Science, Royal Military College of Canada, PO Box 17000 Kingston, ON K7K 7B4, Canada}

\title{MOBSTER: Establishing a Picture of Magnetic Massive Stars as a Population}
\headtitle{MOBSTER}

\begin{document}

\maketitle

\begin{abstract}

Magnetic massive and intermediate-mass stars constitute a separate population whose properties are still not fully understood. Increasing the sample of known objects of this type would help answer fundamental questions regarding the origins and characteristics of their magnetic fields. The MOBSTER Collaboration seeks to identify candidate magnetic A, B and O stars and explore the incidence and origins of photometric rotational modulation using high-precision photometry from the Transiting Exoplanet Survey Satellite (\textit{TESS}) mission. In this contribution, we present an overview of our methods and planned targeted spectropolarimetric follow-up surveys.


\end{abstract}

\section{Introduction}

Recent spectropolarimetric surveys (e.g. MiMeS, BOB, BinaMiCS and LIFE; \citealt{2016MNRAS.456....2W, 2015IAUS..307..342M, 2015IAUS..307..330A, 2018MNRAS.475.1521M}) have revealed the presence of strong, globally organized magnetic fields on the surfaces of $\lesssim$ 10\% of massive and intermediate-mass stars. These magnetic OBA stars form a distinct subpopulation, and neither their incidence rate (which is puzzlingly flat across a large range of stellar masses) nor the apparent lack of correlation between their magnetic and stellar properties are yet fully understood. Their fields are found to be stable over decades (e.g. \citealt{2012MNRAS.419..959O, 2018MNRAS.475.5144S, 2019MNRAS.483.3127S}). These characteristics are fundamentally at odds with the expected properties of dynamo-generated fields; furthermore, while such fields could be generated in the cores of hot stars, models fail to explain how they could be transported to the surface (e.g. \citealt{2001ApJ...559.1094C}). Therefore, the current consensus is that massive and intermediate-mass stars host \textit{fossil} fields \citep{1982ARA&A..20..191B}, i.e. remnants from an earlier stage of evolution, though there are competing theories as to what that stage might be (e.g. pre-main sequence evolution, \citealt{2017arXiv170510650A, 2019A&A...622A..72V}; or a previous stellar merger, \citealt{2019Natur.574..211S}).\footnote{\citet{2015IAUS..305...61N} provide a useful review of the various hypotheses for the origin of magnetic fields at the surfaces of OBA stars.} 

Many open questions remain, not only regarding the uncertain origin of these fields, but also regarding their strengths: the dearth of magnetic fields with strengths between a few G and about 100 G has led to the hypothesis of the existence of a so-called ``magnetic desert'' \citep{2007A&A...475.1053A}, and the evolution of that strength over time is a matter of some debate. Indeed, given the high conductivity in the atmosphere of these stars, magnetic flux conservation is generally assumed. While this might be a valid assumption for A-type stars (e.g. \citealt{2007A&A...470..685L, 2019MNRAS.483.3127S}), it might not be the case for more massive stars \citep{2008A&A...481..465L, 2016A&A...592A..84F, 2019MNRAS.490..274S}.

One of the main barriers to address these questions is the fairly sparse sample of known magnetic massive and intermediate-mass stars. Since spectropolarimetry is a photon-hungry technique, the most efficient approach to increase the sample size is to rely on indirect diagnostics to select highly probable magnetic candidates and build a targeted spectropolarimetric follow-up survey.

In particular, we aim to leverage the massive amount of data produced by the Transiting Exoplanet Survey Satellite (\textit{TESS}; \citealt{2015JATIS...1a4003R}). Due to its large coverage of the sky ($\sim$85\%), high cadence (either 2 or 30 minutes) and exquisite sensitivity and precision, \textit{TESS} represents an ideal instrument for the study of stellar astrophysics, and in particular magnetic stars.


\section{Rotational modulation}

Magnetic massive and intermediate-mass stars present a rather homogeneous phenomenology in broadband optical photometry. Their light curves are typically characterized by periodic brightness variations that are modulated by the star's rotation period (i.e. rotational modulation). This is caused by two separate physical mechanisms. In the case of O- and early B-type stars, the stellar wind interacts with the magnetic field, which channels and confines it around the magnetic equator to form a dense \textit{magnetosphere} (e.g. \citealt{2002ApJ...576..413U}). As a result, the column density varies with the angle at which the magnetosphere is viewed. Since the magnetic and rotational axes are not generally aligned, this viewing angle changes over the course of a rotational period, as described in the Oblique Rotator Model (ORM; \citealt{1950MNRAS.110..395S}). Therefore, more or less continuum light is scattered out of the line of sight by the magnetosphere as the star rotates, leading to measurable photometric variations (e.g. \citealt{1977ApJ...216L..31H, 2011MNRAS.416.3160W}), which can then be modelled \citep{2015MNRAS.451.2015O, 2019MNRAS.tmp.2573M} using simple parametrizations of the structure of the magnetosphere \citep{2005MNRAS.357..251T, 2016MNRAS.462.3830O}. Additionally, variable wind blanketing due to the magnetic field and the non-spherically symmetric surface mass flux can also contribute to light curve variations on a rotational timescale \citep{2016A&A...594A..75K}.

In the case of cooler stars (later B- and A-type stars), the magnetic field affects atomic diffusion processes in the atmosphere \citep{2010A&A...516A..53A}, leading to chemical inhomegeneities on the stellar surface, which in turn cause brightness spots due to flux redistribution (e.g. \citealt{1973ApJ...179..527M, 2007A&A...470.1089K}). As the star rotates, the spots fall in and out of view, leading to periodic photometric variations (e.g. \citealt{1970ApJ...161..685P}).

\begin{figure}
\includegraphics[width=\textwidth]{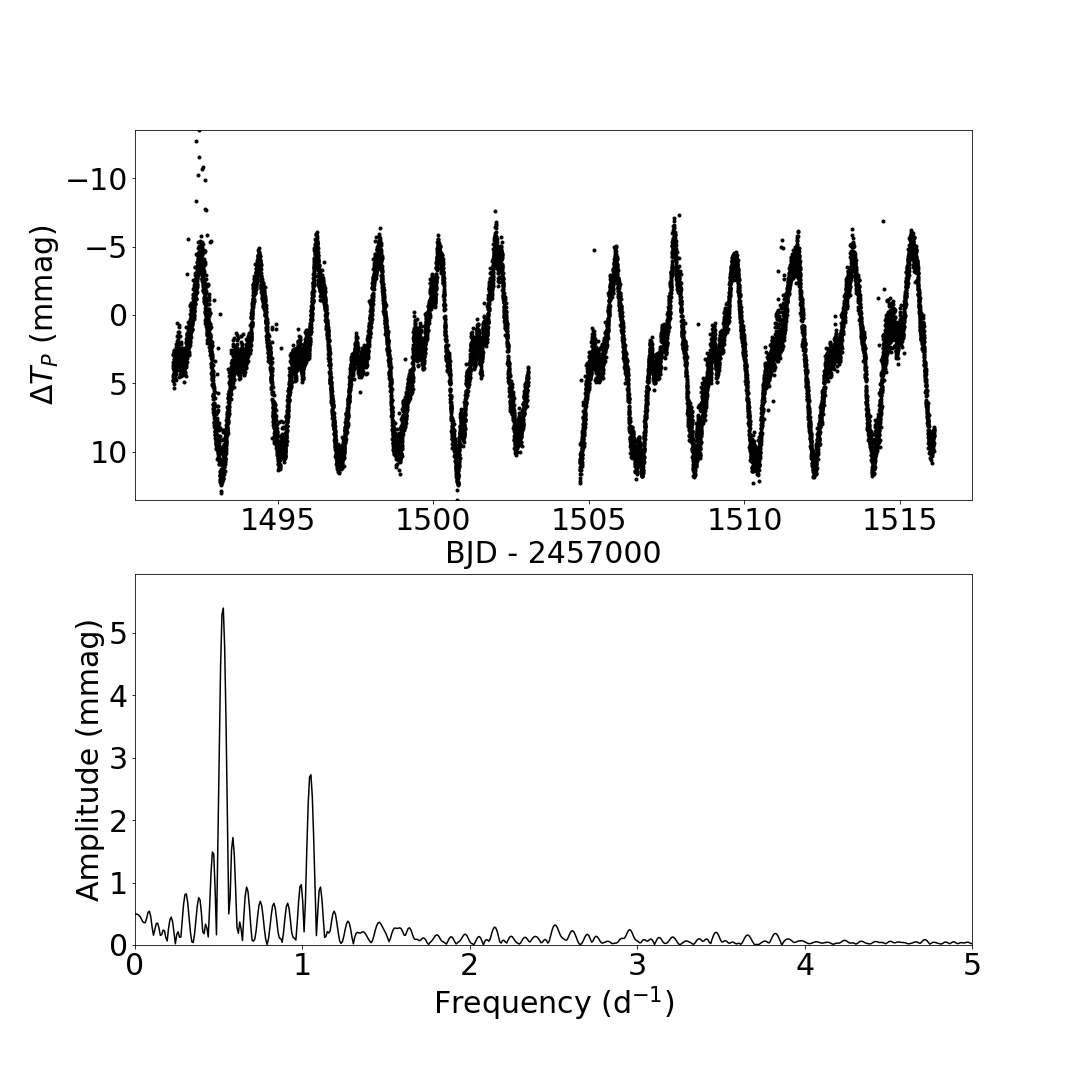}
\caption{Example light curve (top) of the known magnetic B5V star HR 2949 and its corresponding periodogram (bottom), computed using the \texttt{astropy.stats.LombScargle} class in \textsc{python}. Despite the presence of some obvious outliers and some minor instrumental effects in the pipeline processed data (including Pre-search Data Conditioning; e.g. \citealt{2016SPIE.9913E..3EJ}), the main rotational frequency (corresponding to a known period of $\sim$1.9 d, as determined from magnetic data) and its first harmonic are clearly recovered in the periodogram.}
\label{fig1}
\end{figure}

In both cases, the photometric signature is clearly periodic, and can be phased over fairly long time scales. This is already seen in \textit{TESS} data \citep{2019MNRAS.487..304D}, as most of the known magnetic B- and A-type stars that were observed in sectors 1 and 2 exhibited that phenomenology. It was also found in a number of first-light data of B \citep{2019ApJ...872L...9P, 2019MNRAS.485.3457B} and Ap \citep{2019MNRAS.487.3523C} stars. Furthermore, this behavior has been searched for extensively in \textit{Kepler} data \citep{2013MNRAS.431.2240B, 2016MNRAS.457.3724B, 2017MNRAS.467.1830B}, revealing that a substantial fraction of A and B stars might exhibit rotationally-modulated light curves. Pulsations can also lead to similar photometric variations; however, since rotational modulation generally does not manifest itself as a pure sine wave (as expected, given the underlying processes), it can often be diagnosed by the presence of at least one strong harmonic of the rotational frequency (\citealt{2018A&A...616A..77B, 2019MNRAS.487.4695S}; see Fig.~\ref{fig1}). Of course, such a signal could still be associated with other types of variability. Binary stars can undergo eclipses as well as ellipsoidal variations, which are also periodic. While the first case leads to a rather recognizable signature, the second one can be quite easily confused with rotational modulation. Therefore, while an initial light curve analysis based on the detection of a strong, well-defined base frequency and its first harmonic can yield rotational modulation candidates, that sample will need to be further refined.

Since a star cannot be rotating faster than a certain critical velocity, an upper limit can be placed on a putative rotational frequency, restricting the parameter space somewhat. Then, a dedicated literature search can help eliminate other options (e.g. lack of large radial velocity variations that could be ascribed to a binary companion, etc.). Once obvious imposters have been removed, the remaining sample can reasonably be assumed to exhibit rotational modulation with a high probability (although further observations can help confirm that inference). 

Detecting rotational modulation does not necessarily imply the presence of a magnetic field, but it is generally considered as a good indirect diagnostic, as not many physical processes are expected to lead to this type of signature. Therefore, stars which have rotationally modulated light curves can be considered as candidate magnetic stars, but require further diagnostics to confirm the presence of a field.





\section{Spectropolarimetric follow-up}

The most robust and informative observational diagnostic of the presence of a stellar magnetic field is high-resolution spectropolarimetry. Zeeman splitting can be measured in circularly polarized spectra and multi-line techniques can be leveraged to increase signal-to-noise ratios (SNR) and improve sensitivity \citep{1997MNRAS.291..658D}. Existing large surveys are essentially magnitude limited: for instance, the sample studied by MiMeS is largely restricted to stars with V magnitudes brighter than about 6 \citep{2016MNRAS.456....2W}. Addressing our issue with small number statistics will involve surveying fainter stars. However, this poses some practical challenges: fainter stars are much more numerous and obtaining good SNR becomes increasingly challenging. Given that spectropolarimetry is inherently expensive, blind surveys are no longer a viable strategy in order to efficiently improve the number of detected magnetic stars in that magnitude range.

Instead, targeted surveys can strongly increase the efficiency of such an observing program. Using prior knowledge to pre-select a sample with a high probability of detection requires indirect diagnostics of stellar magnetism. As established in the previous section, rotationally-modulated light curves provide one such diagnostic. This method has already proven to be successful: out of a subset of 16 stars selected from the K2 sample for which spectropolarimetric observations were obtained, \citet{2018MNRAS.478.2777B} detected a magnetic field in 11 stars, in other words achieving a detection rate of $\sim$70\%, which is of course much higher than the overall incidence rate ($\sim$10\%) of magnetic fields in hot stars. This strategy does not allow us to assess or refine the global incidence rate, 
but it maximizes the output of a given allocation of telescope time and improves the statistics of magnetic properties.

The three workhorse instruments presently accessible to our community to perform high-resolution spectropolarimetry are NARVAL on the T\'{e}lescope Bernard-Lyot, ESPaDOnS on the Canada-France-Hawaii Telescope, and HARPSpol on ESO's 3.6-m telescope at La Silla. A few principal investigator (PI) proposals have already been submitted to use these facilities 
on a first set of targets. 
Moving forward, our strategy is to compile a comprehensive list of high-probability candidate magnetic stars and 
observe them systematically, taking full advantage of all three instruments. Given their latitudes, they nicely cover the area of the sky probed by \textit{TESS}. At this point, we have identified a few hundred stars whose light curves show signs of rotational modulation. Over the full duration of the mission this number will increase, but we aim to distill this sample to 100-200 high probability candidates, distributed across a range of mass and spectral type bins. Even taking into account a conservative detection rate of 50\%, we expect to usefully increase the number of known magnetic massive and intermediate-mass stars and outperform blind surveys by a factor of a few.

Finally, for even fainter stars, we can attempt to use the FORS2 low-resolution spectropolarimeter on the ESO Very Large Telescope. 
Although this instrument is less sensitive in terms of the detectable magnetic geometries because of its low spectral resolution, it would allow us to explore a much larger sample. Medium-resolution observations can also be carried out using the dimaPol and MSS (Main Stellar Spectrograph) spectropolarimeters, installed respectively on the Dominion Astrophysical Observatory's 1.8-m Plaskett Telescope and the Bolshoi Teleskop Altazimutalny at the Russian Special Astrophysical Observatory.



\section{Conclusions and future work}

The MOBSTER Collaboration is a wide scale community effort comprised of observers and theorists using \textit{TESS} data to improve our knowledge of magnetic massive stars. A large part of that endeavor involves identifying magnetic candidates by selecting stars with rotationally-modulated light curves and building a large targeted spectropolarimetric follow-up survey. By detecting tens to hundreds of new magnetic stars in the OBA spectral type range, we will crucially improve our sample statistics, thus refining the properties of the subpopulation of magnetic massive and intermediate-mass stars and allowing us to answer some of the big questions that have perplexed researchers in our field for the past decade.

Further improvements to our methods should include an automated algorithm to classify light curves and select for rotational modulation. Such a technique has been used fruitfully in the past, sometimes in combination with machine learning (e.g. \citealt{2011MNRAS.414.2602D}). So far, we have mostly been using 2-minute cadence data from \textit{TESS}. However, only about 200,000 targets (including a small fraction of OBA stars) have been observed with this cadence. The full-frame image data (30-minute cadence) includes over 500 million point sources \citep{2018AJ....156..102S}, and will present a much greater challenge in terms of data management and analysis.

Finally, a better theoretical characterization of the expected variability for the most massive stars in our sample would be beneficial to better understand how to recover the rotational modulation signal. Depending on the geometry and the density of the magnetosphere, the photometric variations associated with continuum scattering can be quite subtle (of order a few mmag). However, 
if the magnetospheric material is not centrifugally supported (as is usually the case for magnetic O stars), it falls back onto the stellar surface, forming dynamic flows. Since the magnetosphere is not in a steady state, this can also lead to rather stochastic photometric variations, the amplitude of which has yet to be predicted. This is akin to the line profile variations that are expected (and might be seen) in H$\alpha$ due to this same phenomenon \citep{2013MNRAS.428.2723U}.

\acknowledgements{ADU and GAW acknowledge support from the Natural Sciences and Engineering Research Council of Canada (NSERC). Some of the research leading to these results has received funding from the European Research Council (ERC) under the European Unions Horizon 2020 research and innovation programme (grant agreement No. 670519: MAMSIE). CE acknowledges graduate assistant salary support from the Bartol Research Institute in the Department of Physics, University of Delaware, as well as support from program HST-GO-13629.002-A that was provided by NASA through a grant from the Space Telescope Science Institute. MES acknowledges financial support from the Annie Jump Cannon Fellowship, endowed by the Mount Cuba Observatory and supported by the University of Delaware. AuD acknowledges support from NASA through Chandra Award number
TM7-18001X issued by the Chandra X-ray Observatory Center, which is
operated by the Smithsonian Astrophysical Observatory for and on behalf
of NASA under contract NAS8-03060.}

\bibliographystyle{ptapap}
\bibliography{refs}

\begin{thebibliography}{47}
\providecommand{\natexlab}[1]{#1}
\providecommand{\url}[1]{\texttt{#1}}
\providecommand{\urlprefix}{URL }
\providecommand{\eprint}[2][]{\url{#2}}

\bibitem[{{Alecian} et~al.(2015)}]{2015IAUS..307..330A}
{Alecian}, E., et~al., in G.~{Meynet}, C.~{Georgy}, J.~{Groh}, P.~{Stee} (eds.)
  New Windows on Massive Stars, \emph{IAU Symposium}, volume 307, 330--335
  (2015)

\bibitem[{{Alecian} et~al.(2017)}]{2017arXiv170510650A}
{Alecian}, E., et~al., \emph{arXiv e-prints} arXiv:1705.10650 (2017)

\bibitem[{{Alecian} \& {Stift}(2010)}]{2010A&A...516A..53A}
{Alecian}, G., {Stift}, M.~J., \emph{\aap} \textbf{516}, A53 (2010)

\bibitem[{{Auri{\`e}re} et~al.(2007)}]{2007A&A...475.1053A}
{Auri{\`e}re}, M., et~al., \emph{\aap} \textbf{475}, 3, 1053 (2007)

\bibitem[{{Balona}(2013)}]{2013MNRAS.431.2240B}
{Balona}, L.~A., \emph{\mnras} \textbf{431}, 3, 2240 (2013)

\bibitem[{{Balona}(2016)}]{2016MNRAS.457.3724B}
{Balona}, L.~A., \emph{\mnras} \textbf{457}, 4, 3724 (2016)

\bibitem[{{Balona}(2017)}]{2017MNRAS.467.1830B}
{Balona}, L.~A., \emph{\mnras} \textbf{467}, 2, 1830 (2017)

\bibitem[{{Balona} et~al.(2019)}]{2019MNRAS.485.3457B}
{Balona}, L.~A., et~al., \emph{\mnras} \textbf{485}, 3, 3457 (2019)

\bibitem[{{Borra} et~al.(1982){Borra}, {Landstreet}, \&
  {Mestel}}]{1982ARA&A..20..191B}
{Borra}, E.~F., {Landstreet}, J.~D., {Mestel}, L., \emph{\araa} \textbf{20},
  191 (1982)

\bibitem[{{Bowman} et~al.(2018)}]{2018A&A...616A..77B}
{Bowman}, D.~M., et~al., \emph{\aap} \textbf{616}, A77 (2018)

\bibitem[{{Buysschaert} et~al.(2018)}]{2018MNRAS.478.2777B}
{Buysschaert}, B., et~al., \emph{\mnras} \textbf{478}, 2, 2777 (2018)

\bibitem[{{Charbonneau} \& {MacGregor}(2001)}]{2001ApJ...559.1094C}
{Charbonneau}, P., {MacGregor}, K.~B., \emph{\apj} \textbf{559}, 2, 1094 (2001)

\bibitem[{{Cunha} et~al.(2019)}]{2019MNRAS.487.3523C}
{Cunha}, M.~S., et~al., \emph{\mnras} \textbf{487}, 3, 3523 (2019)

\bibitem[{{David-Uraz} et~al.(2019)}]{2019MNRAS.487..304D}
{David-Uraz}, A., et~al., \emph{\mnras} \textbf{487}, 1, 304 (2019)

\bibitem[{{Donati} et~al.(1997)}]{1997MNRAS.291..658D}
{Donati}, J.~F., et~al., \emph{\mnras} \textbf{291}, 4, 658 (1997)

\bibitem[{{Dubath} et~al.(2011)}]{2011MNRAS.414.2602D}
{Dubath}, P., et~al., \emph{\mnras} \textbf{414}, 3, 2602 (2011)

\bibitem[{{Fossati} et~al.(2016)}]{2016A&A...592A..84F}
{Fossati}, L., et~al., \emph{\aap} \textbf{592}, A84 (2016)

\bibitem[{{Hesser} et~al.(1977){Hesser}, {Ugarte}, \&
  {Moreno}}]{1977ApJ...216L..31H}
{Hesser}, J.~E., {Ugarte}, P.~P., {Moreno}, H., \emph{\apjl} \textbf{216}, L31
  (1977)

\bibitem[{{Jenkins} et~al.(2016)}]{2016SPIE.9913E..3EJ}
{Jenkins}, J.~M., et~al., {The TESS science processing operations center},
  \emph{Society of Photo-Optical Instrumentation Engineers (SPIE) Conference
  Series}, volume 9913, 99133E (2016)

\bibitem[{{Krti{\v{c}}ka}(2016)}]{2016A&A...594A..75K}
{Krti{\v{c}}ka}, J., \emph{\aap} \textbf{594}, A75 (2016)

\bibitem[{{Krti{\v{c}}ka} et~al.(2007){Krti{\v{c}}ka}, {Mikul{\'a}{\v{s}}ek},
  {Zverko}, \& {{\v{Z}}i{\v{z}}{\'n}ovsk{\'y}}}]{2007A&A...470.1089K}
{Krti{\v{c}}ka}, J., {Mikul{\'a}{\v{s}}ek}, Z., {Zverko}, J.,
  {{\v{Z}}i{\v{z}}{\'n}ovsk{\'y}}, J., \emph{\aap} \textbf{470}, 3, 1089 (2007)

\bibitem[{{Landstreet} et~al.(2007)}]{2007A&A...470..685L}
{Landstreet}, J.~D., et~al., \emph{\aap} \textbf{470}, 2, 685 (2007)

\bibitem[{{Landstreet} et~al.(2008)}]{2008A&A...481..465L}
{Landstreet}, J.~D., et~al., \emph{\aap} \textbf{481}, 2, 465 (2008)

\bibitem[{{Martin} et~al.(2018)}]{2018MNRAS.475.1521M}
{Martin}, A.~J., et~al., \emph{\mnras} \textbf{475}, 2, 1521 (2018)

\bibitem[{{Molnar}(1973)}]{1973ApJ...179..527M}
{Molnar}, M.~R., \emph{\apj} \textbf{179}, 527 (1973)

\bibitem[{{Morel} et~al.(2015)}]{2015IAUS..307..342M}
{Morel}, T., et~al., in G.~{Meynet}, C.~{Georgy}, J.~{Groh}, P.~{Stee} (eds.)
  New Windows on Massive Stars, \emph{IAU Symposium}, volume 307, 342--347
  (2015)

\bibitem[{{Munoz} et~al.(2019)}]{2019MNRAS.tmp.2573M}
{Munoz}, M.~S., et~al., \emph{\mnras} 2573 (2019)

\bibitem[{{Neiner} et~al.(2015)}]{2015IAUS..305...61N}
{Neiner}, C., et~al., in K.~N. {Nagendra}, S.~{Bagnulo}, R.~{Centeno},
  M.~{Jes{\'u}s Mart{\'\i}nez Gonz{\'a}lez} (eds.) Polarimetry, \emph{IAU
  Symposium}, volume 305, 61--66 (2015)

\bibitem[{{Oksala} et~al.(2012)}]{2012MNRAS.419..959O}
{Oksala}, M.~E., et~al., \emph{\mnras} \textbf{419}, 2, 959 (2012)

\bibitem[{{Oksala} et~al.(2015)}]{2015MNRAS.451.2015O}
{Oksala}, M.~E., et~al., \emph{\mnras} \textbf{451}, 2, 2015 (2015)

\bibitem[{{Owocki} et~al.(2016)}]{2016MNRAS.462.3830O}
{Owocki}, S.~P., et~al., \emph{\mnras} \textbf{462}, 4, 3830 (2016)

\bibitem[{{Pedersen} et~al.(2019)}]{2019ApJ...872L...9P}
{Pedersen}, M.~G., et~al., \emph{\apjl} \textbf{872}, 1, L9 (2019)

\bibitem[{{Peterson}(1970)}]{1970ApJ...161..685P}
{Peterson}, D.~M., \emph{\apj} \textbf{161}, 685 (1970)

\bibitem[{{Ricker} et~al.(2015)}]{2015JATIS...1a4003R}
{Ricker}, G.~R., et~al., \emph{Journal of Astronomical Telescopes, Instruments,
  and Systems} \textbf{1}, 014003 (2015)

\bibitem[{{Schneider} et~al.(2019)}]{2019Natur.574..211S}
{Schneider}, F. R.~N., et~al., \emph{\nat} \textbf{574}, 7777, 211 (2019)

\bibitem[{{Shultz} et~al.(2018)}]{2018MNRAS.475.5144S}
{Shultz}, M.~E., et~al., \emph{\mnras} \textbf{475}, 4, 5144 (2018)

\bibitem[{{Shultz} et~al.(2019)}]{2019MNRAS.490..274S}
{Shultz}, M.~E., et~al., \emph{\mnras} \textbf{490}, 1, 274 (2019)

\bibitem[{{Sikora} et~al.(2019{\natexlab{a}}){Sikora}, {Wade}, {Power}, \&
  {Neiner}}]{2019MNRAS.483.3127S}
{Sikora}, J., {Wade}, G.~A., {Power}, J., {Neiner}, C., \emph{\mnras}
  \textbf{483}, 3, 3127 (2019{\natexlab{a}})

\bibitem[{{Sikora} et~al.(2019{\natexlab{b}})}]{2019MNRAS.487.4695S}
{Sikora}, J., et~al., \emph{\mnras} \textbf{487}, 4, 4695 (2019{\natexlab{b}})

\bibitem[{{Stassun} et~al.(2018)}]{2018AJ....156..102S}
{Stassun}, K.~G., et~al., \emph{\aj} \textbf{156}, 3, 102 (2018)

\bibitem[{{Stibbs}(1950)}]{1950MNRAS.110..395S}
{Stibbs}, D.~W.~N., \emph{\mnras} \textbf{110}, 395 (1950)

\bibitem[{{Townsend} \& {Owocki}(2005)}]{2005MNRAS.357..251T}
{Townsend}, R.~H.~D., {Owocki}, S.~P., \emph{\mnras} \textbf{357}, 1, 251
  (2005)

\bibitem[{{ud-Doula} \& {Owocki}(2002)}]{2002ApJ...576..413U}
{ud-Doula}, A., {Owocki}, S.~P., \emph{\apj} \textbf{576}, 1, 413 (2002)

\bibitem[{{ud-Doula} et~al.(2013)}]{2013MNRAS.428.2723U}
{ud-Doula}, A., et~al., \emph{\mnras} \textbf{428}, 3, 2723 (2013)

\bibitem[{{Villebrun} et~al.(2019)}]{2019A&A...622A..72V}
{Villebrun}, F., et~al., \emph{\aap} \textbf{622}, A72 (2019)

\bibitem[{{Wade} et~al.(2011)}]{2011MNRAS.416.3160W}
{Wade}, G.~A., et~al., \emph{\mnras} \textbf{416}, 4, 3160 (2011)

\bibitem[{{Wade} et~al.(2016)}]{2016MNRAS.456....2W}
{Wade}, G.~A., et~al., \emph{\mnras} \textbf{456}, 1, 2 (2016)

\end{thebibliography}

\end{document}